\documentclass[twocolumn,apl,amsmath,amssymb,showpacs,superscriptaddress]{revtex4-2}
\usepackage{epsf} 
\usepackage{graphicx}
\usepackage{color}
\usepackage{soul}
\usepackage{gensymb}
\usepackage{sidecap}
\usepackage{amsmath}
\usepackage{mathtools}
\usepackage{float}
 \usepackage{multirow}
\usepackage[hidelinks,colorlinks=true,linkcolor=blue,citecolor=blue]{hyperref}
\DeclareUnicodeCharacter{2212}{\textendash}

\begin{document}
\title{Flexible Trilayer Cellulosic Paper Separators engineered with BaTiO$_3$ ferroelectric fillers for High Energy Density Sodium-ion Batteries}
\author{Simranjot K. Sapra}
\affiliation{Department of Physics, Indian Institute of Technology Delhi, Hauz Khas, New Delhi-110016, India}
\affiliation{International College of Semiconductor Technology, National Yang Ming Chiao Tung University, 1001 University Road, Hsinchu 30010, Taiwan}
\affiliation{Department of Materials Science and Engineering, National Yang Ming Chiao Tung University, 1001 University Road, Hsinchu 30010, Taiwan}
\author{Mononita Das}
\affiliation{Energy Materials and Devices Division, CSIR-Central Glass and Ceramic Research Institute, Kolkata- 700032, India} 
\author{M. Wasim Raja}
\affiliation{Energy Materials and Devices Division, CSIR-Central Glass and Ceramic Research Institute, Kolkata- 700032, India} 
\author{Jeng-Kuei Chang}
\affiliation{International College of Semiconductor Technology, National Yang Ming Chiao Tung University, 1001 University Road, Hsinchu 30010, Taiwan}
\affiliation{Department of Materials Science and Engineering, National Yang Ming Chiao Tung University, 1001 University Road, Hsinchu 30010, Taiwan}
\affiliation{Department of Chemical Engineering, Chung Yuan Christian University, 200 Chung Pei Road, Taoyuan, 32023 Taiwan}
\author{Rajendra S. Dhaka}
\email{rsdhaka@physics.iitd.ac.in}
\affiliation{Department of Physics, Indian Institute of Technology Delhi, Hauz Khas, New Delhi-110016, India}

	\date{\today}      
	
	\begin{abstract}
Cellulose-based paper separators are employed in sodium-ion batteries (SIBs) as a viable and economical substitute of conventional separators, owing to their sustainability, scalability, safety and cost-effectiveness. We design a full cell configuration having Na$_{3}$V$_{2}$(PO$_{4}$)$_{3}$ as cathode and pre-sodiated hard carbon as an anode with different separators and compare the electrochemical performance of these ceramic-impregnated polymer-coated cellulose paper separators with commercial glass fiber separator. Notably, the paper-based multilayer separators provide desirable characteristics such as excellent electrolyte wettability, thermal stability up to 200\degree C, and ionic conductivity, which are essential for the efficient operation of SIBs. The cellulose separator is coated by a layer of polyvinylidene fluoride polymer, followed by a second layer of styrene butadiene rubber (SBR) polymer in which ferroelectric fillers BaTiO$_{3}$ are integrated, which interacts with the polymer hosts through Lewis acid-base interactions ion and improves the conduction mechanism for the Na$^{+}$ ions. The final lamination is performed by varying the SBR concentrations (0.5, 0.75, and 1.0 w/v\%). The incorporated polymer matrices improve the flexibility, adhesion and dispersion of the nanoparticles and affinity of the electrolyte to the electrode. The morphology of the paper separators shows the uniform interconnected fibers with the porous structure. Interestingly, we find that the paper separator with 0.75 w/v\% content of SBR exhibit decreased interfacial resistance and improved electrochemical performance, having retention of 62\% and nearly 100\% Coulombic efficiency up to 240 cycles, as compared to other concentrations. Moreover, we observe the energy density around 376 Wh kg$^{-1}$ (considering cathode weight), which found to be comparable to the commercially available glass fiber separator. Our results demonstrate the potential of these multilayer paper separators towards achieving sustainability and safety in energy storage systems. 


	\end{abstract}
	\maketitle

	\section{\noindent ~Introduction}
	
The abundance of sodium resources on Earth has been the driving force behind the emergence of sodium-ion batteries (SIBs) as efficient and ecologically friendly energy storage technologies, with the objective of achieving sustainable and green power storage system \cite{Hwang_CSR_2017, Roberts_NSA_2018, Deng_AEM_2018, Rudola_JMCA_2021}. In the interim, SIBs continue to captivate researchers as potential replacements for lithium-ion batteries (LIBs) due to their cost-effectiveness and comparable intercalation mechanism \cite{Deng_AEM_2018, Rudola_JMCA_2021, Sapra_Wiley_2021, Manish_CEJ_2023, Dwivedi_AEM_2021, Pati_JMCA_2022}. Among the primary components of a battery, the function of a separator is to prevent a short circuit by creating a physical barrier between the electrodes and ensuring that the battery's functionality is improved by the presence of sufficient pores that facilitate the passage of ions during the charge-discharge process \cite{Xue_NanoToday_2024, Wu_SEF}. The shutoff mechanism of the separators is essential for maintaining the thermal stability, i.e, in the event of battery overheating, it melts and halts the flow of ions, thereby averting thermal runaway \cite{NiuACS-AEM21, ZhangESM21}. The separators also demonstrate great electrolyte wettability, which is a critical factor in enhancing ionic conductivity and further optimizing the battery performance \cite{Luo_JPS_2021}. The introduction of electric vehicles has significantly altered the performance requirements of separators, despite their long-standing use in the industry \cite{Pan_EES_2013}. Consequently, it is imperative to create separators that can withstand the difficult operational conditions by exhibiting remarkable thermal and mechanical stability. Currently, a variety of separators are employed in battery applications, including polyolefin materials such as polyethylene (PE) and polypropylene (PP) \cite{Babiker_JPS_2023, Xue_NanoToday_2024, Wang_epolymer_2019}. They are easy to fabricate, have high porosity, and are optimized in thickness, exhibit excellent mechanical and chemical stability. Nevertheless, these polyolefin membranes exhibit inadequate thermal stability, which presents significant safety hazards. This is due to the fact that they experience severe thermal contraction at elevated temperatures as a result of their low melting points.

Another drawback of PE separators is their natural surface hydrophobic behaviour, which results in poor wettability in liquid electrolytes \cite{Babiker_JPS_2023}. The other class of commercially available glass fibers comprise the non-metallic fibers with a melting point above 500$^{o}$C and good fire resistance; however, poor mechanical strength and high-cost limit their effective utilization. To overcome these challenges, the cellulose-based paper separators have garnered significant attention as alternatives of commercial separators \cite{Mononita_ACSOmega_2023, Das_JPS_2024, Basu_patent}. Importantly, cellulose is the most abundant biomass on earth, readily available, and follows cost-effective manufacturing procedures \cite{Zhou_JPS_2023, Waqas_Small_2019}. Moreover, the cellulose-based paper separators exhibit excellent electrochemical properties in LIBs owing to their great electrolyte compatibility and structural, thermal and mechanical stability \cite{Mononita_ACSOmega_2023, Das_JPS_2024}. They also offer pore size tuning to enhance electrolyte absorption through pore engineering and provide isolation for both electrodes \cite{Lizundia_CPTA_2020}. In addition, there are many ongoing optimization strategies to improve the battery performance and stability \cite{Zhu_AEM_2024}. These separators are engineered with specific functional groups on their surface and layering techniques to overcome the electrolyte decomposition and formation of dendrites \cite{Chen_CEJ_2023}. In this line, the class of separators have emerged based on the number of layers and the addition of active and inert ceramic to tune the properties for the safety of separators \cite{JiangACS-AMI17}. It is also reported that Sumitomo separators used by Tesla and Panasonic have a coating of ceramic particles and aromatic polyamide to increase the penetration strength of ions \cite{Blomgren_JES_2016}. Further, the cellulose separators can be coated with the polyvinylidene fluoride polymer matrix layer to reduce the pore sizes and enhance the surface smoothness \cite{Raja_AEM_2022}. However, the PVDF alone cannot meet the requirement of the safety due to their high shrinkage and swelling/dissolution issues in the non-aqueous liquid electrolyte, which leads to the exfoliating of the coating layer from the separators \cite{Shi_JPS_2014}. Therefore, another layer, comprising styrene butadiene rubber (SBR) polymer, along with the incorporation of ceramic oxide nanoparticles, is coated on it. The SBR polymer is a strong adhesion agent and provides a strong dispersion medium, flexibility and high heat resistance \cite{Buqa_JPS_2006}. 

Interestingly, the Na$^{+}$-free inorganic fillers, BaTiO$_{3}$ are chosen, which possess high melting temperatures and affinities and promote the formation of percolating channels through the Lewis acid-base interactions between the fillers and polymer hosts \cite{Sharma_JES_2021, Das_JPS_2024}. Also, incorporating inorganic ceramic nanoparticles decreases the crystallinity of polymer matrices and provides additional pathways for the transport of Na$^{+}$ ions, thereby enhancing the ionic conductivity of the polymer matrix \cite{Itoh_SSI_2003, Itoh_JPS_2003, Sun_JES_2000}. It increases the wettability of the electrolyte as nanoparticles have high surface area and hydrophilicity. It further improves the mechanical stability to withstand the volume changes upon battery cycling \cite{Arun_AMI_2023, Chen_AFM_2024}. However, the homogenous distribution of ceramic particles, surface charge and the type of deposition of ceramic particles on the surface of the cellulose fibers play significant roles in the cellulose-ceramic-based separator’s optimal performance. To further improve the flexibility, safety and adhesion of the separators in the stack, another layer of SBR polymer is coated on the paper separator with the different concentrations of SBR (0.5, 0.75 and 1.0 w/v\%). However, these types of paper separators have not been explored in sodium-ion batteries for energy storage.  

Therefore, in this study, we investigate the effect of various concentrations of SBR-coated multilayer separators sandwiched in full cell configurations having  Na$_{3}$V$_{2}$(PO$_{4}$)$_{3}$ (NVP) as cathode and pre-sodiated hard carbon (HC) as an anode \cite{Stubble_BSuperCap_2024, Chen_Nanoscale_2019}. A detailed investigation is carried out to determine the structure, morphology, mechanical strength and ionic conductivity of the paper separators. The different concentrations of SBR aqueous solution (0.5, 0.75 and 1.0 w/v\%) are coated onto the composite polymer membrane, and the electrochemical properties of 0.75 w/v\% SBR separator are compared with the commercial glass fiber separators.

\section{\noindent ~Experimental Section}
	
\noindent 2.1 \textit{~Synthesis of Na$_{3}$V$_{2}$(PO$_{4}$)$_{3}$ and Hard Carbon}: 
	
	The Na$_{3}$V$_{2}$(PO$_{4}$)$_{3}$ cathode material is prepared by the ball-milling assisted solid-state method \cite{Sapra_AMI_2024}. We use commercial Hard carbon (HC) powder (purchased from Shaldong Gelon LIB Co. Ltd.) as an anode, having particle size of 9 $\mu$m and specific surface area of $\leq$ 5 m$^{2}$/g.
	
\noindent 2.2 \textit{~Synthesis of Trilayer Paper Separators}:
	
We use the commercial cellulose paper (20 $\mu$m), styrene butadiene rubber (SBR) solution and BaTiO$_{3}$ (BTO) nanopowder (particle size $\leq$ 100 nm) are purchased from Merck, Germany. The trilayer paper separator fabrication process comprises three main steps: sizing, ceramic impregnation, and lamination \cite{Mononita_ACSOmega_2023}. Due to the naturally rough and uneven surface of paper, it's essential to enhance its properties with a suitable polymer. Initially, commercial cellulose paper (referred to as PS) undergoes coating with 3.5 w/v \% poly(vinylidene fluoride) (PVDF) using an in-house designed double-decker fabricator unit with in situ heating by a wet-coating method \cite{Mononita_ACSOmega_2023}. This process, termed ``sizing", strengthens the paper and improves its surface consistency by reducing pore spaces, fiber extrication, and moisture absorption. The resulting PVDF-coated paper, known as P35, is then subjected to a second coating with an aqueous blend of SBR (3.5 w/v\%) and BTO nano-ceramic powder (2.0 w/v\%) in the ``ceramic impregnation" step. This step lowers the surface energy, reinforces fibers with ceramic fillers, and further reduces pore spaces. By combining  the sizing and ceramic impregnation steps enhance the wettability of the separator membrane, resulting in P35S35B20. In the final lamination step, the paper is coated with varying concentrations (0.5--1.0 w/v\%) of SBR aqueous solution to create a smooth surface, minimise pore spaces, and act as a protective layer against electrolyte interaction. After lamination, the paper separators are compacted under pressure and labelled based on coating of SBR concentration: as P35S35B20S050 (0.5 w/v\%), P35S35B20S075 (0.75 w/v\%), and P35S35B20S100 (1.0 w/v\%), which are abbreviated as S050, S075 and S100, respectively. The brief schematic showing the trilayer coating of polymers and fillers on the cellulose separator is shown in Fig.~S1 of \cite{SI}. The commercial battery grade glass fiber separators are used from Advantec, Japan. 

\noindent 2.3 \textit{Materials characterization}: 
	
	We use the laboratory-based X-ray diffraction with Cu K$_{\alpha}$ radiation ($\lambda$ = 1.5406~\AA) to investigate the crystal structure of all the components. The vibrational modes were measured with the attenuated total reflection fourier transmission infrared spectroscopy (ATR-FTIR, Thermo Nicolet-IS-50) in the spectral range of 500--4000 cm$^{-1 }$. The field emission scanning electron microscopy (FE-SEM) imaging of the paper separators is performed using the TESCAN (Model No. Magna LMU) to study their morphology.. The air permeability study of the separators was carried out using a Gurley precision instrument (model no. 4110N). A thickness gauge (model no. CD-6'' CSX, Mitutoyo Corporation, Japan) was used to measure the thickness of the fabricated separators. The thermal shrinkage of the paper separators along with the commercial separator was studied using a circular disk of diameter 15 mm. The membrane disks were placed inside a vacuum oven, and the temperature was raised to 200\degree C at an interval of 50 $\degree$C, with a holding time of 20 min at each temperature. A circular disk of 19 mm diameter for each separator is used to measure the electrolyte-soaking ability of fabricated separator membranes. The mechanical properties of the fabricated separators were measured by the universal testing machine (model no. 5500R, INSTRON, UK). The tensile stress and strain of the samples were measured both in machine direction (MD) and transverse direction (TD) with a crosshead speed of 10 mm/min and 1 kN load. 

\noindent 2.4 \textit{ Fabrication of electrodes and full cell assembly}:
	
	A cathode slurry was prepared by mixing 80 wt.\% active powder, 10 wt.\% super P, and 10 wt.\% PVDF (Polyvinylidene fluoride) in N-methyl-2-pyrrolidone solution. This slurry was pasted onto the battery grade Al foil and vacuum-dried at 90\degree C for 8 hrs. The electrode was punched based on the required dimensions of CR2032 and the active mass loading $\mathrm{\sim}$3 - 3.5 mg cm$^{-2}$ was obtained. Similarly, negative electrode slurry was prepared by mixing 80 wt.\% powder of hard carbon, 10 wt.\% super P, and 10 wt.\% PVDF in N-methyl-2-pyrrolidone solvent. This slurry was pasted onto the battery-grade Cu foil and vacuum-dried at 90\degree C for 8 hrs. The pre-sodiation of the hard carbon against the Na metal anode is conducted at 100 mA/g for 2 cycles to compensate for the sodium loss as HC anode consume sodium to form solid electrolyte interface. After pre-sodiation, the cells were disassembled, and the negative electrode is taken out. Then, full cells were assembled in an argon-filled glove box (UniLab Pro SP from MBraun, Germany) using NVP as the positive electrode and pre-sodiated HC as the negative electrode, maintaining anode-to-cathode capacity ratio between 1.1--1.2. The 1M NaClO$_{4}$ in ethylene carbonate (EC) and polypropylene carbonate (PC) electrolyte with the 10 wt.\% fluoroethylene carbonate (FEC) additive solvent and trilayer membranes are employed as the electrolyte and separators. The glove box maintains moisture and oxygen content $\mathrm{\sim}$0.1 ppm. The cell balance is called the NP ratio, defined by equation~1 \cite{Song_JES_2020}:
\begin{equation}
\frac{N}{P} = \frac{Anode\:Capacity\:(mAh)}{Cathode\:Capacity\:(mAh)}
\label{NP}
\end{equation}
where, N is the anode capacity for the first sodiation and P is the cathode capacity for the first de-sodiation and NP ratio represents the sodium storage capacity of the anode to cathode. The practical capacities in equation \ref{NP} are calculated from the half-cell performance of each electrode against the Na metal \cite{Song_JES_2020}. 
	
\noindent 2.5 \textit{Electrochemical measurements}: 

The charge and discharge testing of the fabricated cells were evaluated between 2.0 and 4.3 V with various currents using Neware BTS400 battery tester at room temperature. The cyclic voltammetry (CV) was studied at sweep rates of 0.05--1.0~mV s$^{-1}$, the electrochemical impedance spectroscopy (EIS) measurements were conducted in the frequency range of 100 kHz to 10 mHz with an AC amplitude of 10 mV, and the bulk resistance of the electrolyte-soaked paper separators was evaluated in the frequency range of 1 mHz to 1 MHz at room temperature using the VMP-3 potentiostat (Biologic Instruments).

\section{\noindent ~Results and discussion}
\begin{figure}
    \centering
    \includegraphics[width=\linewidth]{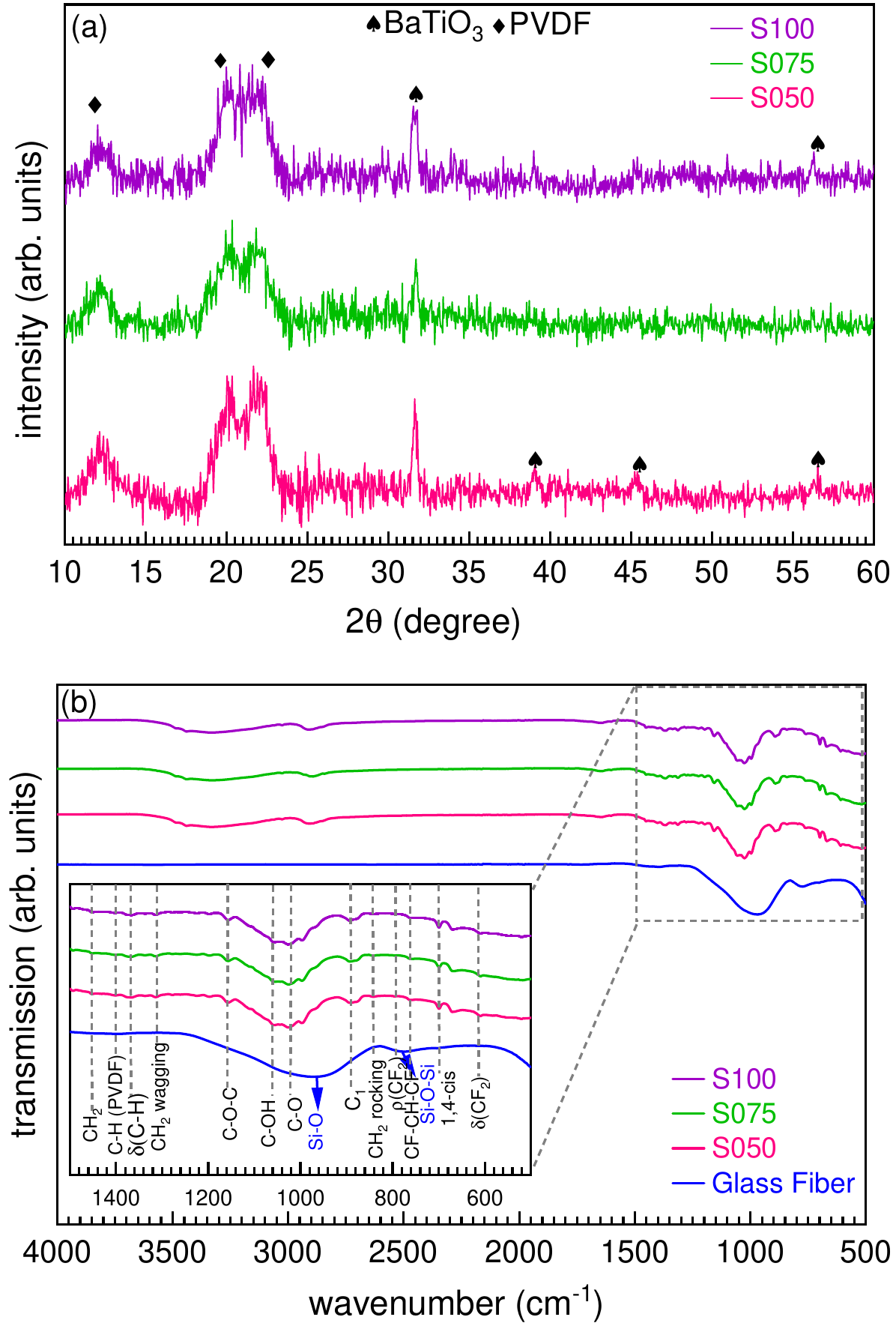}
    \caption{(a) The x-ray diffraction patterns of the S050, S075 and S100 paper separators, and (b) the ATR-FTIR spectra for all the separators (glass fiber, S050, S075 and S100).}
    \label{XRD}
\end{figure}

 \noindent 3.1 \textit{Physical Characterizations}:
 
The X-ray diffraction (XRD) patterns of paper separators are shown in Fig.~\ref{XRD}, which indicate the presence of PVDF and BTO filler in these developed membranes. The diffraction peaks at 2$\theta$ values of 12$\degree$, 20.1$\degree$ and 22$\degree$ correspond to the PVDF polymer and show the crystalline nature. The diffraction peaks observed at 2$\theta$ value of 31.6, 39, 45.3 and 56.7 $\degree$ matches well with the crystalline BTO particles and correspond to the (110), (111), (200) and (220) lattice planes \cite{Arun_AMI_2023}. Hence, the XRD patterns confirm the co-existence of PVDF and BTO crystalline phases in the developed paper separators, which are crucial for understanding the ion transport mechanism. Also, the XRD pattern of the commercial glass fiber (GBR) separator, which comprises of silicon dioxide, is shown in Fig.~S2 of \cite{SI}. We observe a broad peak located at around 24.5$\degree$, which confirms the amorphous state of the glass fiber separators \cite{Luo_JMCA_2018}.

To further investigate the chemical bonding of ceramic-filled polymer membranes, the ATR-FTIR measurements are performed, which has been proven as an effective tool for the assessment of the bonding structure. The spectrum is recorded in the range of 500--4000 cm$^{-1}$ for the glass fiber and paper separators, as shown in Fig.~\ref{XRD}(b). The most significant identification of the cellulose is provided by the spectral band at 1030 cm$^{-1}$ for the cellulose anhydroglucose C--O vibrations, whereas spectral bands at 1313 and 1369 cm$^{-1}$ are associated with the CH$_{2}$ wagging and acetyl C--H deformation vibrations of cellulose \cite{Wolfs_JPS_2023}. The absorption bands at 892.4, 1108.8, 1156.6, and 1451.2 cm$^{-1}$ are ascribed to group C1 frequency, ring asymmetric stretching, C--O--C asymmetric stretching, and CH$_{2}$ symmetric bending vibration in cellulose, respectively. The spectral bands at 1060 cm$^{-1}$ are associated with --C--OH groups in narrower band in cellulose \cite{Iglesias_BioMacromolecules_2019}. The band at 844 and 764 cm$^{-1}$ can be assigned to the CH$_{2}$ and CF$_{2}$ rocking and skeletal bending of C(F)--C(H)--C(F) of the PVDF for the different polymorphs, which arises from the macromolecular chains \cite{Kubin_PT_2023}. The spectral band at 1400 cm$^{-1}$ is ascribed to the cellulose's stretching vibrations (--CH bonding). The spectral band at  612 and 794 cm$^{-1}$ belongs to $\delta$(CF$_{2}$) and $\rho$(CF$_{2}$) vibrations of the $\alpha$ phase of PVDF \cite{Sharma_JES_2021}. The band observed at 698.6 cm$^{-1}$ belongs to the 1,4-cis mode of the SBR polymer \cite{Lopez_ePolymer_2005}. The spectral bands in the range 2900--3600 cm$^{-1}$ are correspond to the stretching vibrations of OH and hydrogen bonds between the oxygen containing functional groups and water, while peak at 1640 cm$^{-1}$ is attributed to the stretching vibrations of the structural water \cite{Casas_ACSAMI_2020, Li_RSC_2023}. Interestingly, we observe that adding BaTiO$_{3}$ fillers in the SBR polymer matrix does not cause a significant band shifting and the ATR--FTIR spectra are dominated by the characteristic features of cellulose. 

\begin{figure*}
    \includegraphics[width=0.95\linewidth]{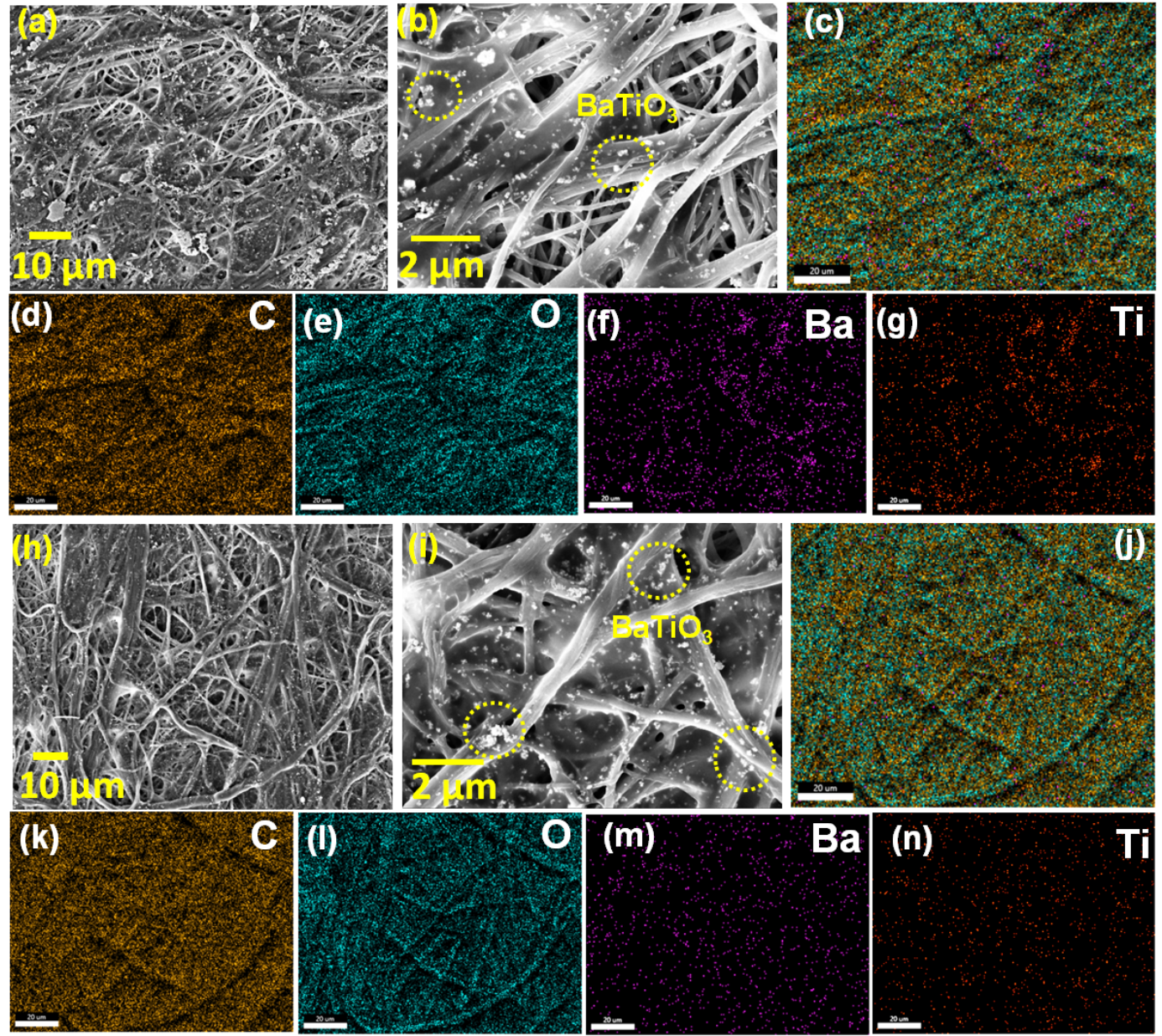}
    \caption{The FE-SEM images for the S075 paper separator and the corresponding elemental mappings in the selected regions in (a) and (h) for the top surface (a--g) and bottom surface (h--n), respectively.}
    \label{fesem}
\end{figure*}

\begin{figure}
    \centering
    \includegraphics[width=\linewidth]{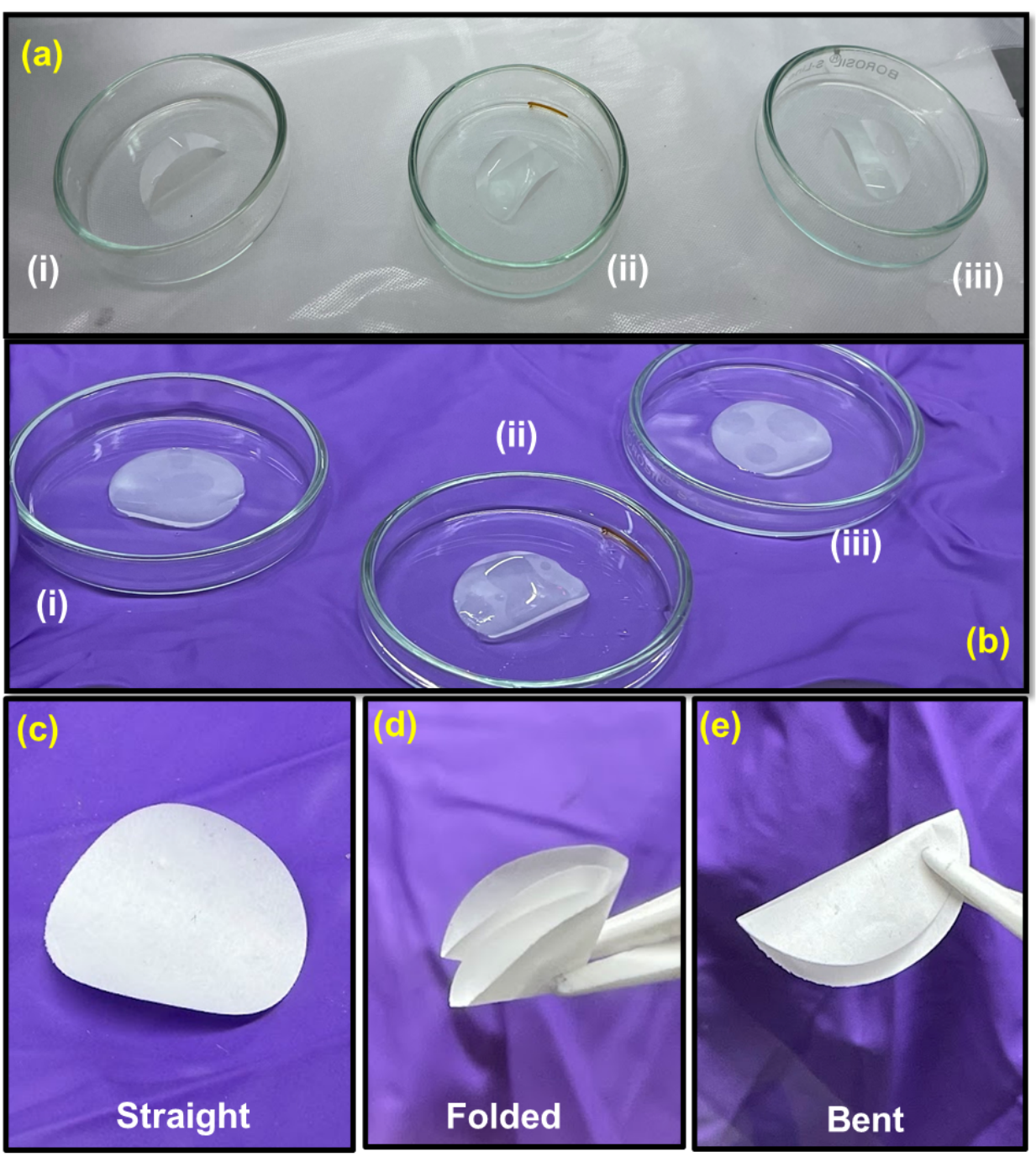}
    \caption{The electrolyte soaking ability test for the paper separators S050 (i), S075 (ii) and S100 (iii), soaked in the liquid electrolyte for (a) 0 hrs and (b) 2 hrs; (c-e) the digital photographs of the paper separator, S075 at various stages of folding and bending.}
    \label{electrolyte}
\end{figure}

\begin{table*}
     \caption{The comparison of the thickness, air permeability, pore characteristics, and Young's Modulus along with the MD and TD direction of the developed paper separators.}
    \centering
    \begin{tabular}{p{1.5cm} p{1.5cm} p{2.5cm} p{2.3cm} p{3cm} p{3cm} p{3.5cm}}
    \hline
    separator&thickness&air permeability &surface area &total pore volume &Pore diameter&Young modulus MD/TD \\
     &$\mu$m& (Gurley in s)&(m$^{2}$ g$^{-1}$)& (cc g$^{-1}$)&(nm)& (MPa)\\
\hline
S050&20&20.3&5.503&5.303$\times$10$^{-3}$&3.058&2897.48/1057.02\\
S075&20&13.0&10.523&1.425$\times$10$^{-2}$&3.280&1926.71/745.37\\
S100&20&17.2&7.121&8.714$\times$10$^{-3}$&3.314&2938.61/937.25\\	
 \hline
\hline
  \end{tabular}
    \label{physical}
\end{table*}

  \begin{figure*}
  \centering
    \includegraphics[width=1\linewidth]{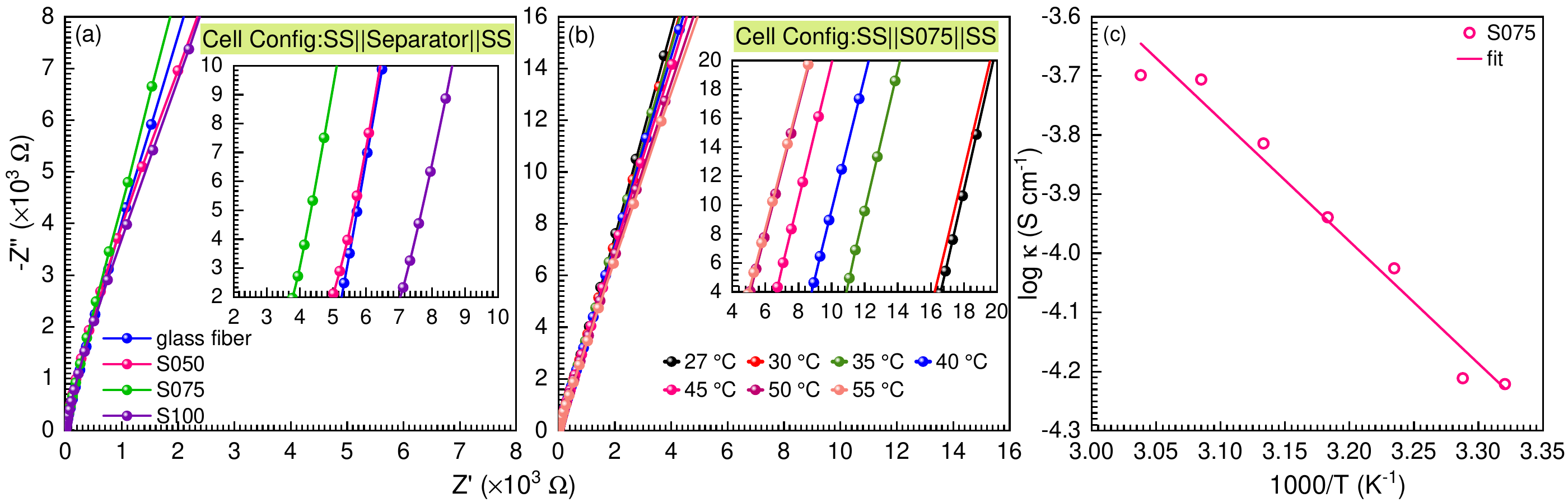}
    \caption{(a) The Nyquist plot for all the separators at the room temperature; (b) the Nyquist plot for the control, S075 membrane at the various temperatures (27\degree C, 30\degree C, 35\degree C, 40\degree C, 45\degree C, 50\degree C, 55\degree C), using a blocking electrode system (SS$\parallel$Separator$\parallel$SS); (c) the ionic conductivity values for the S075 separator at the different temperatures.}
    \label{Bulk conductivity}
\end{figure*}

The microstructure of the paper separator, S075 and glass fiber separator (GBR) is studied using the FE-SEM measurements, as shown in Fig.~\ref{fesem} and Fig.~S3 of \cite{SI}, respectively. Figs.~\ref{fesem}(a) and (b) depict the top surface of the BaTiO$_3$ filled polymer coated cellulose separator, where the cellulose fibers are entangled and nanosized BaTiO$_{3}$ particles are embedded inside the polymer matrix along with the surface. It is observed that the ceramic impregnation along with the lamination of the SBR polymer reduced the porosity and increased the compact density. The uniform distribution of the cellulose fibers and ferroelectric nanofillers are depicted in Figs.~\ref{fesem}(c--g). Similarly, the bottom side of the paper separator shows the successful impregnation of the BaTiO$_{3}$ nanoparticles in the entangled cellulose fibers, depicted in Figs.~\ref{fesem}(h--n). The BaTiO$_{3}$ particles lying on the surface of the fibers increases compact density, which is highly expected to increase the mechanical strength of the cellulose separators. The microstructure of the commercial glass fibers consisting of the nano-sized fibers, is shown in Figs.~S3(a--e) of \cite{SI} with the uniform distribution of C, O and Si elements.

The electrolyte soaking capability experiment for the prepared paper separators is conducted inside the glovebox, where 50$\mu$L of standard electrolyte 1 M NaClO$_{4}$ in EC: PC (1:1) with the 10 wt\% FEC additive is dropped on each of the separators and allowed to soak for about 2 hrs. The weight of the dry (4.3 mg) and wet separators (10.8--11.1 mg) is recorded, and electrolyte uptake is quantified as per the equation~2, where W$_{\rm dry}$ and W$_{\rm wet}$ are the weight of the separators before and after soaking the liquid electrolyte. We have also attached the images of the separators before and after soaking in the electrolyte, as shown in Figs.~\ref{electrolyte}(a-b). The liquid uptake of the electrolyte is calculated as 152--158\%, which is close to the reported values in refs.~\cite{Mononita_ACSOmega_2023, Das_JPS_2024}. Also, we demonstrate the flexibility of the paper separators with the folding and bending which is crucial to withstand the stress, as shown by the digital pictures in Figs.~\ref{electrolyte}(c-f).
\begin{equation}
{\rm electrolyte-uptake} = \frac{W_{\rm wet} - W_{\rm dry}}{W_{\rm dry}}\times 100\%
\label{electrolyte}
\end{equation}   
The Gurley value serves as an essential physical indicator for evaluating the air permeability of porous membranes, with particular relevance in characterising separators for battery applications \cite{Jang_Materials_2020}. In this study, a highly porous cellulosic paper substrate (PS) undergoes modification by integrating duo-polymers (PVDF and SBR) as binders and BaTiO$_{3}$ ceramic nanoparticles as fillers. It is anticipated that introducing these materials may influence the air permeability values of the separator membranes if pore spaces within the paper substrate are obstructed either by the polymer binders or by the formation of large aggregates from nano-sized BTO particles. The Gurley value of PS is measured to be 5 s. After sizing with 3.5 w/v\% PVDF, the value increases marginally to 8 s, indicating that the low concentration of polymer has minimal impact on overall porosity. With the incorporation of ceramics on PVDF-coated paper via ceramic impregnation process and lamination with SBR layer, there is only a little increase in Gurley values: 20.3 s for S050, 13.0 s for S075, and 17.2 s for S100. The lower Gurley values of all the prepared paper separators demonstrate the larger size of pore spaces, indicating higher air permeability, which means facilitating smooth Na-ion transport across the separator \cite{Mononita_ACSOmega_2023}. The Gurley value for all the paper separators is shown as bar chart in Fig.~S6(a) of \cite{SI}. 

The stress and strain values are recorded in the machine direction (MD) and transverse direction (TD) to evaluate the tensile strength of the paper separators. It is noteworthy that the tensile strength of any paper or paper-modified matrix, such as paper separators, inherently relies on the strength of cellulosic tissue, their cross-linking nature, and inter- and intra-hydrogen bonding among cellulose units or added functional polymers \cite{Wohlert_Cellulose_2022, Francolini_IJMS_2023}. In this case, the inclusion of polymeric binders (PVDF and SBR) in paper separators plays a crucial role by forming additional hydrogen bonds with cellulosic hydroxyl groups and interaction of the polymer with the ceramic filler BTO further adds to the significant enhancement in the mechanical strength (46--50 MPa in MD and 25--28 MPa in TD) \cite{Mononita_ACSOmega_2023}. The stress strain curves to evaluate the tensile strength of separators is shown in Fig.~S6(b) of \cite{SI}. Furthermore, it has been observed that the developed paper separators retain their structural integrity at elevated temperatures when heated up to 200\degree C for 30 mins. The digital photographs showing the thermal stability of the paper separators at 25$\degree$C and 200$\degree$C are shown in Fig.~S7 of \cite{SI}. This demonstrates the improved dimensional stability of the paper separators at elevated temperatures, which prevents any deformation and guarantees their safe use in sodium-ion batteries. The surface area, average pore diameter and total pore volume of the developed separators are summarized in Table~\ref{physical}. The physical characteristics of the NVP and HC electrodes in the half-cell configurations are reported in refs.~\cite{Sapra_AMI_2024, Madhav_IJPAP_2024}.  \\

\noindent 3.2 \textit{Impedance Spectroscopy}: \\
The electrochemical impedance spectroscopy (EIS) measurements were conducted in symmetric full cell configuration with paper separators using the ion-blocking method. The separators were immersed in a liquid electrolyte solution of 1M NaClO$_{4}$ in EC and PC (1:1) ratio for about 2 hrs. They were then placed between two stainless steel (SS) electrodes with a diameter of 16 mm and tightly sealed in a 2032 coin cell configuration within a glovebox. The EIS data are represented through Nyquist plots in Fig.~\ref{Bulk conductivity}, which show straight-line profiles inclined towards the Z$^{'}$ axis for all the separators, signifying the characteristic behavior of electrode-electrolyte double-layer capacitance. However, there is no semicircle region, which means the major contribution comes from the diffusion of Na--ions from the electrolyte salt, represented by the straight line and the major charge carriers are ions. The resistance values of the paper separators are calculated by analysing the high-frequency intercept of the Nyquist plot on the Z$^{'}$ axis \cite{Mononita_ACSOmega_2023} and are provided in Table~\ref{Bulk Conductivity}. 
\begin{table}
 \caption{The resistance and bulk conductivity values for all the developed paper separators and temperature-dependent conductivity values for the S075 separator.}
  \begin{tabular}{p{2.5cm} p{2.2cm} p{3.09cm}}
  \hline
        composition&resistance&ionic conductivity\\
		&($\Omega$)&(S cm$^{-1}$)\\
\hline
         Glass Fiber&5.2&1.9 $\times$ 10$^{-4}$\\
         S050&4.8& 2.1 $\times$ 10$^{-4}$\\
         S075 &3.7&2.7 $\times$ 10$^{-4}$ \\
         S100 &7.0&1.4 $\times$ 10$^{-4}$ \\
   		\hline
\hline
   		temperature&resistance&ionic conductivity\\
		(\degree C)&($\Omega$)&(S cm$^{-1}$)\\
		\hline
		27&16.6&0.60 $\times$ 10$^{-4}$\\
          30&16.2&0.61 $\times$ 10$^{-4}$\\
          35&10.5&0.94 $\times$ 10$^{-4}$ \\
          40 &8.6&1.15 $\times$ 10$^{-4}$ \\
	     45&6.5&1.53 $\times$ 10$^{-4}$\\
          50&5.1&1.97 $\times$ 10$^{-4}$ \\
          55&5.0&2.00 $\times$ 10$^{-4}$ \\
\hline
\hline
    \end{tabular}
    \label{Bulk Conductivity}
\end{table}
From the bulk resistance values, we calculate the ionic conductivity of the paper separators using equation \ref{eqn-con}: 
\begin{equation}
\kappa = L/R_{b}A
\label{eqn-con}
\end{equation}
where L, A, R$_{b}$ are the thickness, area of the contact of separators and bulk resistance, respectively. The ionic conductivity is found to be highest for the S075 (2.7 $\times$ 10$^{-4}$) as compared to others, confirming its more electrolyte soaking capability and providing easier ion migration pathways for Na--ion transportation through the electrostatic interactions of the polymer with the high dielectric constant and ferroelectric BTO particles. Also, the microporous structure formed by the SBR, PVDF and BTO nanoparticles could improve the electrolyte absorption so that Na--ions can be transferred sufficiently and smoothly into the separator. However, when the SBR content reaches 1.0 w/v\%, the conductivity decreases indicating that the pore blockage arising from the complex polymer ceramic interactions with increasing the SBR content. To understand the transport phenomena, the ionic conductivity of the control separator, S075, is calculated from the impedance spectra at different temperatures. The Nyquist plots of S075 separator at the various temperatures are shown in Fig.~\ref{Bulk conductivity}(b). As the ionic conductivity depend on the temperature, the bulk resistance varies and follows the Arrhenius model as per the equation~\ref{arrhenius}:
\begin{equation}
\kappa = \kappa_{o}e^{\frac{-E_a}{RT}}
\label{arrhenius}
\end{equation}
where R, $\kappa_{o}$ and E$_{a}$ are the gas constant (8.314 J mol$^{-1}$ K$^{-1}$), pre-exponential factor and apparent activation energy, respectively \cite{Barbosa_AMI_2023}. The activation energy value, calculated from the slope in Fig.~\ref{Bulk conductivity}(c), is found to be 1.8 $\times$10$^{-4}$ J/mol. Here, we observe the linear trend of the ionic conductivity with the temperature, which indicates the increased mobility of ions and expansion of polymeric chains leading to an increase in the free volume, which enhances the segmental motion of the polymer chains and Na ion transport \cite{Rohan_JMCA_2015}. Hence, the modified coating approach enhances the ability of the paper separator to interact with the electrolyte, resulting in decrease the energy barriers for ion migration at the interface. \\

\begin{figure*}
    \includegraphics[width=1.0\linewidth]{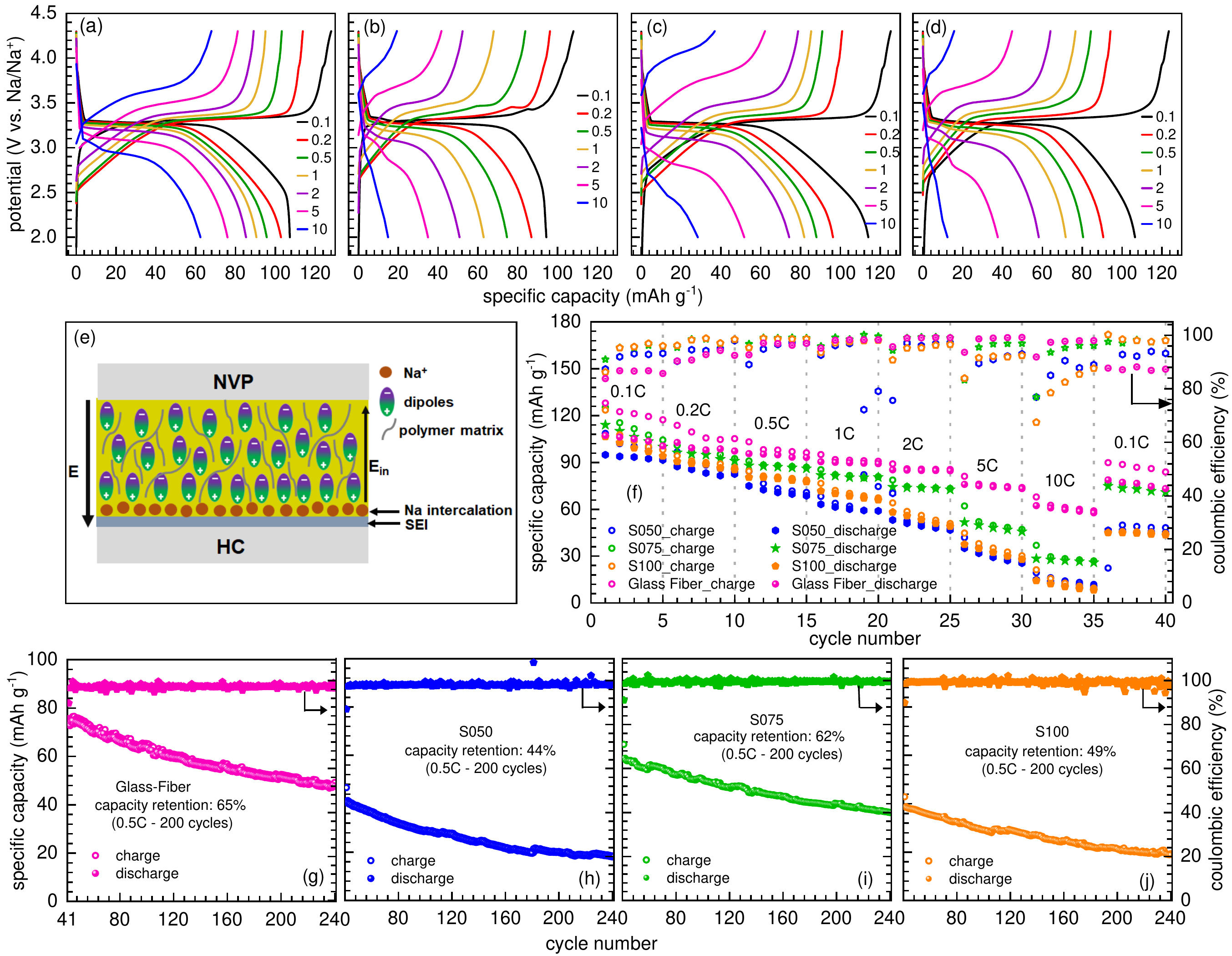}
    \caption{The electrochemical performance of the S050, S075, S100 and glass fiber separator based full cell system: (a--d) the galvanostatic charge-discharge profiles; (e) the mechanism for the functioning of separators; (f) the rate performance at the different C rates; (g-j) the cycle life and Coulombic efficiency at 0.5 C rate between 41 and 240 cycles (cycle life is measured after the rate performance is finished in 40 cycles for the same set of cells).}
    \label{GCD}
\end{figure*}

\noindent 3.3 \textit{~Electrochemical performance of full cells}\\
\noindent 3.3.1 \textit{~Cell performance}:\\
In order to investigate the sodium charge storage mechanism and compare between paper separators and commercial glass fiber separators, we test the electrochemical performance in full cell using NVP$\parallel$Separator$\parallel$HC configuration at room temperature. The electrochemical performance of HC and NVP electrodes against Na metal in half cell configurations is discussed in detail in Fig.~S4 of \cite{SI} and \cite{Sapra_AMI_2024}, respectively. The pre-sodiation of HC anode is performed to compensate for the irreversible Na loss before the full cell fabrication. The separator membranes are soaked in the electrolyte for 2 hrs before the cell assembly, and the electrode balance is maintained between 1.1 and 1.2. After that, the assembled full cells are cycled at lower current densities (here 0.1 C, 1 C = 117.6 mA g$^{-1}$ based on theoretical capacity of cathode) to form stable solid electrolyte interface (SEI), allowing the smooth migration of Na ions across the electrode-electrolyte interface. The galvanostatic charge-discharge (GCD) profiles for the full cells are depicted in Figs.~\ref{GCD}(a--d) from 0.1 C to 10 C current rates for commercial glass separator and paper separators S050, S075, S100, respectively. The GCD curves show a flat plateau at 3.3 V vs Na/Na$^{+}$, corresponding to the V$^{3+}$/V$^{4+}$ redox couple in the NVP cathode \cite{Jian_AM_2017}. The initial discharge capacities at 0.1 C rate are 95, 114 and 106.5 mAh g$^{-1}$ (based on cathode only) in case of the S050, S075, and S100, respectively, which are comparable to that of glass fiber (107.2 mAh g$^{-1}$), while the S075 shows the highest value. Interestingly, when the full cells are cycled from 0.2 C to 10 C (at different current densities), the high rate performance is observed for all the developed paper separators. This can be attributed to the favourable interfacial charge transport between the electrode and electrolyte, as the SBR coating layer assists in adhering the separator to the electrodes after soaking in the electrolyte solution, thereby confirming the compatibility of the paper separators similar to the glass fiber separators. 

As a general trend, the specific discharge capacity decreases with the increasing C-rate due to the ohmic polarization and interfacial resistances for the ion transfer and diffusion limitation. However, as shown in  Figs.~\ref{GCD}(a--d), the specific capacity values at a higher rate (10 C) for the developed paper separators are less than the commercial glass fiber separators, which may be due to the mechanical interface issues as the interface becomes quite rigid with the incorporation of ceramic BTO particles in the polymer matrix. Also, the presence of SBR coating layers with ceramic fillers may have reduced the free space for ion migration due to pore blockage and non-uniform pore distribution through the separators. The rate performance at the different current rates for every 5 cycles is depicted in Fig.~\ref{GCD}(f), where the S075 and glass fiber show superior rate capability as compared to others. After the initial SEI formation, the cycling stability improved at different current rates. The long-cycle electrochemical performance, along with the Coulombic efficiencies for all the separators up to 240 cycles, is shown in Figs.~\ref{GCD}(g--j). It is observed that with the S075 separator the cell displays a capacity retention of 62\%, which is comparable to the glass fiber (65\%) case; however, with the S050 and S100 separators the cells show stronger capacity fading with retention of 44\% and 49\%. This behavior signify that the SBR concentration in S050 separator is not sufficient enough to laminate the cellulose separator and reduce the pore spaces; while, the excess concentration of SBR in S100 separator may laminate the separator so thicker, so the polymer segregation with the partial pore blockages possibly cause the capacity decay \cite{Mononita_ACSOmega_2023, Das_JPS_2024, Raja_AEM_2022}. On the other hand, the Coulombic eﬃciency in all the full cells is maintained nearly 100\% up to 200 cycles, which proves the effective transport of sodium ions through all the separators for sodium-ion batteries. 

\begin{figure*}
    \centering
    \includegraphics[width=0.95\linewidth]{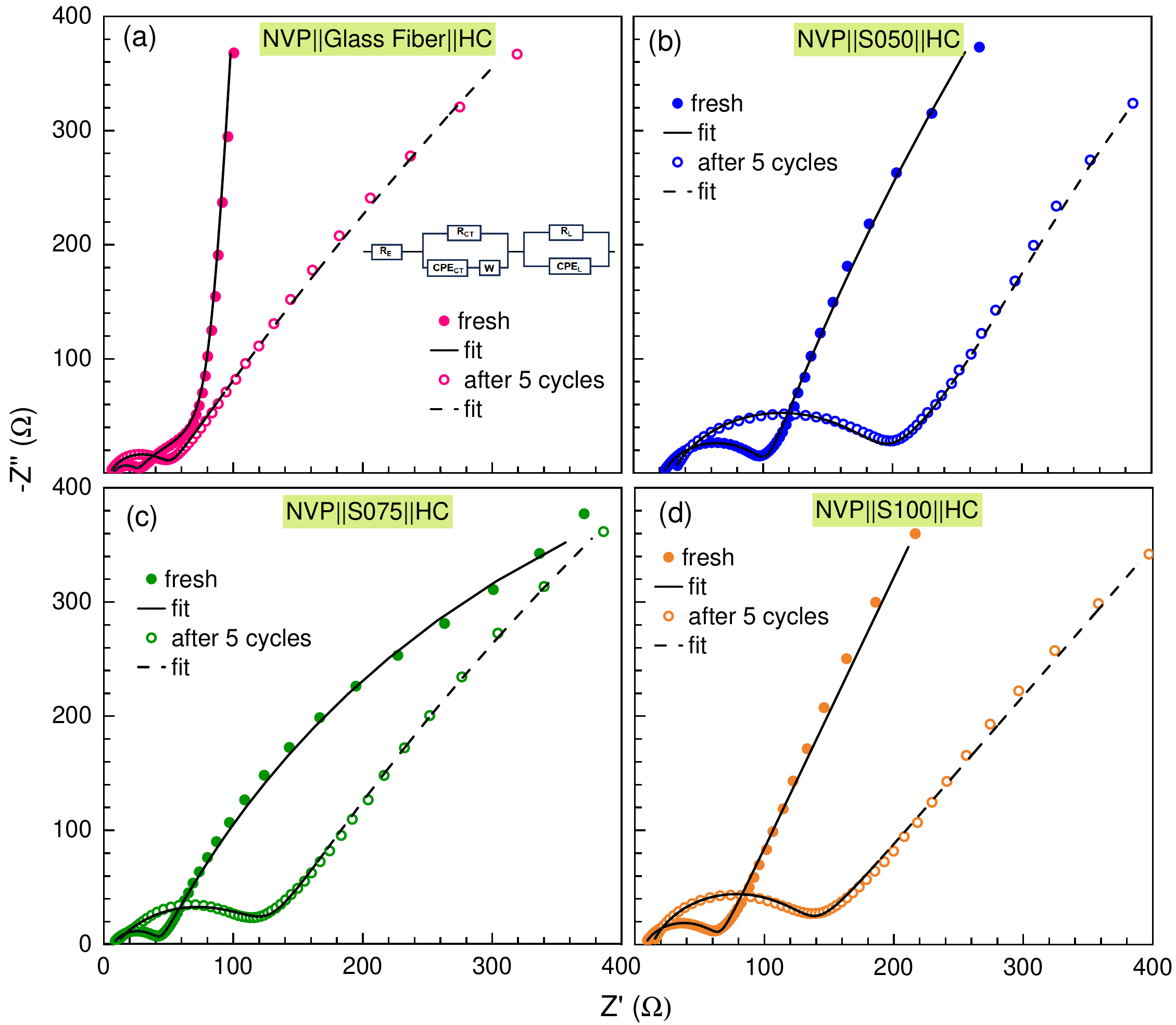}
    \caption{(a--d) The Nyquist plots for the NVP-HC full cells with glass fiber, and paper separator (S050, S075 and S100) for the fresh and after stabilizing the SEI at 0.1 C for five cycles, along with the fitting using Randles circuit shown in the insets.}
    \label{EIS}
\end{figure*}

The enhancement in the electrochemical performance of S075 paper separator in full cell is orchestred by the combined roles of the cellulose, ceramic BTO filler, PVDF and SBR polymers and NaClO$_{4}$ salts in the electrolyte for the favorable sodium ion diffusion and suppress the sodium dendritic growth. The integration of nanosized BTO particles with the polymer matrices through the Lewis-acid-base relationship improves the ionic conduction mechanism with the increase in charge carrier concentration by facilitating salt dissociation. The surface charge of BTO particles is also expected to increase the volume of amorphous phase due to the high nucleation rate of crytsallization process where fillers act as small nucleation agents \cite{Sun_JES_1999, Sun_JES_2000}. In addition, the ferroelectric ceramic filler BTO possess high dielectric constant and permanent dipole moment, where the alignment and reorientation of dipoles is used to tune the ionic transport and polarization. During the charging process, an internal piezoelectric field (E$_{in}$) is introduced, as depicted in Fig.~\ref{GCD}(e). This field is uniformly distributed in the electrolyte and directs ion migration towards the anode with controlled ion transport pathways, leading to uniform intercalation of Na ions and diminishes the growth of dendrites \cite{Dong_NanoL_2024}. The BTO ceramic filler provides mechanical strength to withstand the pressure during cycling and an additional SBR lamination along with the PVDF coating on the separators provides flexibility and offers structural integrity for the ceramic and polymer matrices in the cellulose framework \cite{Mononita_ACSOmega_2023}. 

Nevertheless, the 0.75 w/v\% of SBR content is optimum to ensure sufficient pore spaces with improved wettability and complex polymer ceramic interactions, enhancing Na ion's facile transportation in the host polymer matrices. As the concentration of the SBR further increases, the electrochemical performance starts declining, which can be associated with the partial pore blockage due to the complex polymer and filler interactions and the partial agglomeration of BTO nanoparticles is also possible \cite{Mononita_ACSOmega_2023, Das_JPS_2024}. The other reason associated with the capacity fading can be attributed to the complex interactions of the SBR with the electrode particles as the isolating character of the binder makes poor electrical contact with the particles, which further hinders the sodium ion diffusion during the sodiation/de-sodiation process. Therefore, it is found that 0.75 w/v\% SBR is the optimal concentration of these trilayer cellulose separators for the enhanced electrochemical performance. These observations are also consistent with the similar trend observed in the air permeability (Gurley value) and mechanical properties of prepared paper separators, as discussed above and in \cite{SI}. Furthermore, the full cell performance in the provided potential window (2.0--4.3 V) proves the electrochemical stability of these developed paper separators at high potentials due to the presence of the electron-withdrawing C--F functional groups and high dielectric constant of PVDF polymer \cite{Zhu_SRep_2013}. The prepared multilayer paper separators also have excellent electrolyte retention as ceramic particles enhance the sodium ion penetration across the multilayer of the polymer matrices, which provide complex polymer ceramic interactions and the ease of movement of the Na ions is enhanced through the creation of transport channels. Also, it provides good compatibility with the sodium-based redox couples stability, which show average voltage of 4.0 V (here NVP) and therefore, these can be used in the high voltage SIBs. In addition, the energy density for the full cell battery is calculated (considering the cathode weight) to be 351.5, 270, 376 and 346.5 Wh kg$^{-1}$ for the glass fiber, S050, S075, and S100, respectively. These values confirm the superior performance of battery when S075 separator is used. 

\begin{table}[h]
   \caption{The values of the charge transfer resistance (R$_{ct}$) before and after cycling for 0.1 C rate for the 5 cycles for the interface stabilisation for all the separators.}
  \begin{tabular}{p{2.6cm} p{2.6cm} p{2.6cm}}
\hline
        Composition&pre-cycling ($\Omega$)&post-cycling ($\Omega$)\\
\hline
         Glass Fiber&21.2&43.2\\
         S050&78.0& 164.0\\
         S075 &36.8&117.4 \\
         S100 &37.3&120.3 \\
   		\hline
    \end{tabular}
    \label{cycling}
\end{table}

\noindent 3.3.2 \textit{~Impedance Spectroscopy before and after cycling}: \\
	 Moreover, the EIS technique is employed to probe into the sodium-ion transport across the electrode/electrolyte interface. The EIS data are recorded for the fresh coin cells after 12 hrs of rest period, and later, the cells were cycled at 0.1 C rate for five cycles to allow the formation of a stable electrode/electrolyte interface. The Nyquist plot for pre and post-cycling measurements at room temperature are presented in Figs.~\ref{EIS}(a--d), where the inset in panel (a) shows the equivalent Randles's circuit model. The equivalent Randles's circuit unveils the resistances involved in migrating sodium ions through the electrode-electrolyte interfaces. The intercept of the semicircle demonstrates the bulk resistance (R$_{E}$), and R$_{ct}$  describes the charge transfer resistance across the electrode-electrolyte interface in the high-frequency region. The slopy region in the low-frequency region provides information about the diffusion of Na ions across the bulk of the electrode, corresponding to the Warburg diffusion component, W. The CPE is the constant phase element incorporated for the polycrystalline and rough electrodes, corresponding to the double-layer capacitance at the solid electrolyte interphase, justifying that the interfacial and bulk charge transfer controls diffusion process \cite{Sapra_AMI_2024}. By fitting the given electrochemical system with the mentioned equivalent circuit [inset of Fig.~\ref{EIS}], we find that the interfacial impedance increase with the time and the values are depicted in Table.~\ref{cycling}. 
		 
	 The charge transfer resistance of post-cycling for the S075-based full cell is 117.4 $\Omega$, which is much lower than 164 $\Omega$ for S050, and 120.3 $\Omega$ for S100 separators. However, the glass fiber separators show the value of R$_{ct}$ = 43.2 $\Omega$, which is even much smaller than the S075, indicating reduced interfacial resistance upon cycling and indicating that the stabilization of the passivating layer/SEI layer in glass fiber takes place more easily, whereas, the paper separators suffer the challenge of mechanical interface. Here, the BTO concentration in the developed paper separators can be further optimized to improve the interfacial resistance. 
	\begin{figure}[h]
    \centering
    \includegraphics[width=\linewidth]{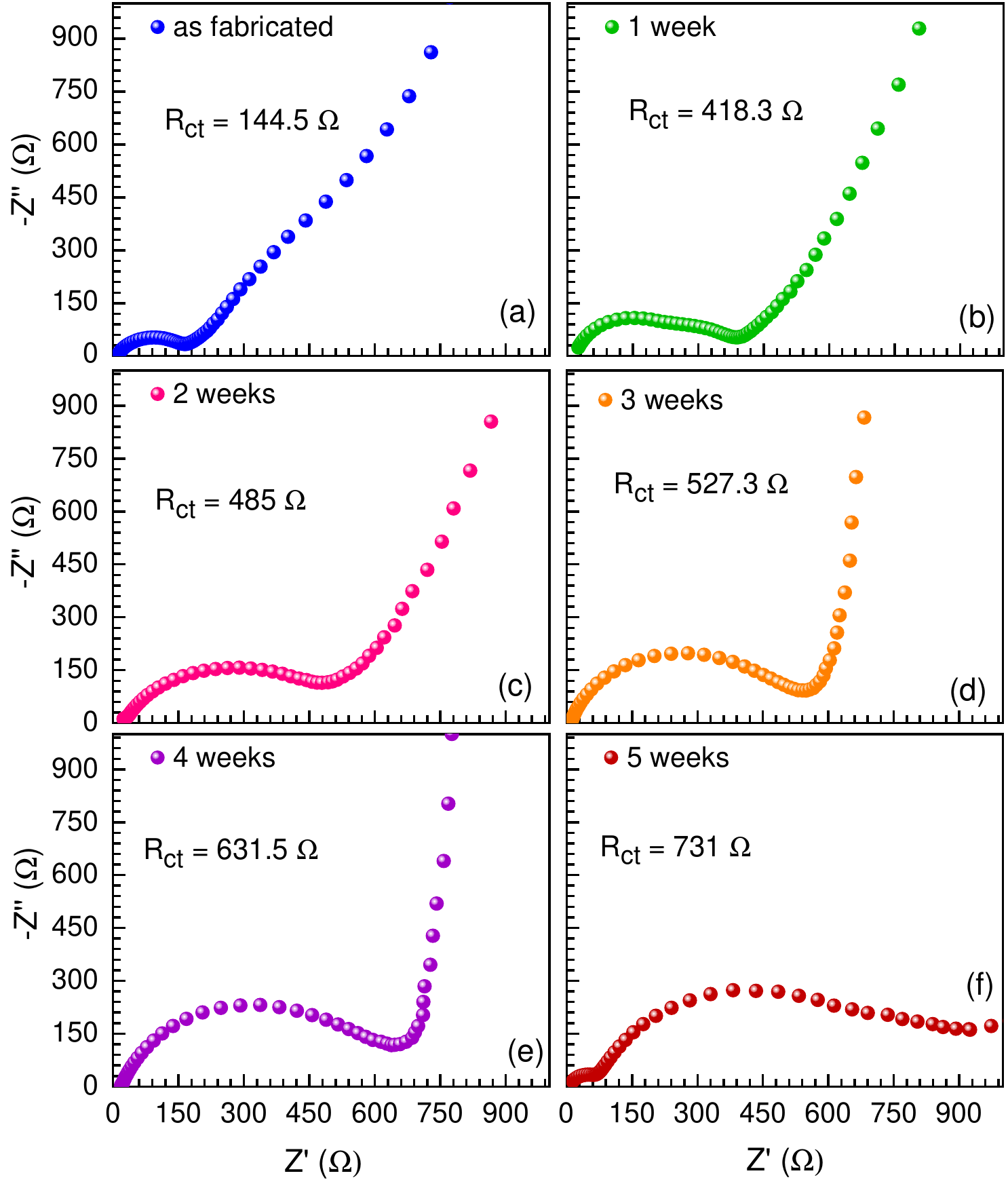}
    \caption{The Nyquist plots for the aged NVP-HC full cells using S075 paper separator for 0, 1, 2, 3, 4 and 5 weeks.}
    \label{EIS-aged}
\end{figure}	 
However, we observe an increasing trend in the interfacial resistance after the cycling for all the paper and commercial separators, which maybe related to the decreased ionic transport arising from the deteriorated interface, and low Coulombic efficiency of the HC anode can also affect the electrochemical cycling, which aligns well with the GCD profiles and rate performance \cite{Mohsin_JPS_2022}. This is because of the irreversible loss of sodium during the growth of SEI layer, as a result, the R$_{ct}$ increases \cite{Arora_JES_1998}. These results signify the similar compatibility of the trilayer paper separators as the glass fiber separators, having the NVP and HC as electrodes, for the easy transportation of ions across the electrode-electrolyte interface and further optimization of the electrolyte and pre-sodiation methods can enhance the electrochemical performance of full cells \cite{Xu_AEM_2018}. \\

\noindent 3.3.3 \textit{Calendar Ageing Test}: \\
Finally, the calendar ageing test is conducted to study the ageing processes that lead to the degradation of the battery performance, independent of the charge and discharge cycling \cite{Krupp_JES_2022, Keil_JES_2016}. Herein, the study of the evolution of passivation layers across the electrode-electrolyte interface is the primary mechanism of calendar ageing, as the formation, growth, or reconstruction of passivation layers consume cyclable sodium as a result of electrolyte decomposition at the cathode interface and formation of the SEI at the anode side \cite{Mohsin_JPS_2022}. Therefore, the impedance evolution across the interface for the full cells with S075 separator is studied through EIS at OCV over several weeks, as the Nyquist plots are presented in Figs~\ref{EIS-aged}(a--f). It is observed that charge transfer resistance increases with the passage of time from 144.5, 418.3, 485, 527.3, 631.5 and 731 $\Omega$ for 0, 1, 2, 3, 4 and 5 weeks, respectively, which could be due to the reactions at the interface with the FEC of the electrolyte which decomposes to form side products. In addition, the large amounts of irreversible sodium loss in the formation of SEI layer as well as cathodic electrolyte interface (CEI) on the HC and NVP interface, respectively, are the significant factors, which lead to increase in the charge transfer resistance in the aged electrodes \cite{Arora_JES_1998, Zheng_JES_1999}. 

\section{\noindent ~Conclusions}

We have successfully prepared the sustainable, cost effective, and thin multilayer cellulose based separator and tested it in full-cell systems designed for sodium-ion batteries using the Na$_{3}$V$_{2}$(PO$_{4}$)$_{3}$ as cathode and hard carbon as an anode. The polymer matrices, impregnated with ceramic (BaTiO$_{3}$), are applied to cellulose-based paper separators. The polymer matrices have varying content of SBR at concentrations of 0.5\%, 0.75\%, and 0.10\% by weight/volume. The paper-based ceramic separators demonstrate excellent electrolyte wettability, thermal stability, and ionic conductivities, which are crucial for the efficient operation of SIBs. The structure of the paper separators was validated using X-ray diffraction and Fourier transform infrared spectroscopy. The presence of BaTiO$_{3}$ nanofillers facilitated the effortless movement of the Na ions, as evidenced by ionic conductivity measurements. The galvanostatic charge-discharge profiles demonstrate enhanced electrochemical efficiency and an excellent capacity retention up to 200 cycles at a rate of 0.5 C for the paper separator with 0.75 w/v\% SBR content. It also depicts lower changes in the interfacial changes after cycling, as compared to other concentrations, as revealed by the pre-- and post--electrochemical impedance spectroscopy. Our research demonstrated comparable electrochemical performance to the commercially available glass fiber separator, indicating that the paper separators has the potential to contribute in the development of efficient and sustainable SIBs for energy storage. \\


	
\section{\noindent ~Acknowledgements}
SKS thanks the IIT Delhi for the fellowship, and Jayashree Pati and Manish K. Singh for help during experiments as well as useful discussion. We thank IIT Delhi for providing the FTIR research facility at the Central Research Facility and the Physics department for the XRD facility. RSD acknowledges SERB--DST for financial support through a core research grant (file no.: CRG/2020/003436). MWR acknowledges the financial support from CSIR, GoI through TAPSUN and FTT projects for establishment of roll to roll fabrication facilities for paper separators. RSD and MWR also acknowledge the Department of Science and Technology (DST), Government of India, for financial support through ``DST--IIT Delhi Energy Storage Platform on Batteries" (project no. DST/TMD/MECSP/2K17/07). RSD and JKC thank NYCU, Taiwan and IIT Delhi for the support through the M-FIRP project (MI02683G).


\end{document}